\documentclass[12pt,a4paper]{article}
\usepackage{fullpage}
\usepackage[centertags]{amsmath}
\usepackage{amsbsy}
\usepackage{amsfonts}
\usepackage{amssymb}
\usepackage{eucal}
\usepackage[dvips]{graphicx,color}
\allowdisplaybreaks[1]
\newcommand{\openone}{\leavevmode\hbox{\small1\kern-3.8pt\normalsize1}}

\newcommand{\vet}[1]{\ensuremath{\hskip-1pt\vec{\hskip1pt#1}}}

\begin{document}
\begin{titlepage}
\begin{flushright}
\begin{tabular}{l}
hep-ph/0102320
\\
DFTT 3/2001
\\
February 27, 2001
\end{tabular}
\end{flushright}
\vspace{1cm}
\begin{center}
\large\bfseries
Lepton Numbers in the framework of Neutrino Mixing
\\[0.5cm]
\normalsize\normalfont
S.M. Bilenky
\\
\small\itshape
Joint Institute for Nuclear Research, Dubna, Russia, and
\\
\small\itshape
INFN, Sez. di Torino, and Dip. di Fisica Teorica,
Univ. di Torino, I--10125 Torino, Italy
\\[0.3cm]
\small\normalfont
and
\\[0.3cm]
\normalsize\normalfont
C. Giunti
\\
\small\itshape
INFN, Sez. di Torino, and Dip. di Fisica Teorica,
Univ. di Torino, I--10125 Torino, Italy
\end{center}
\begin{abstract}
In this short review we discuss the notion of lepton numbers. 
The strong evidence in favor of neutrino oscillations
obtained
recently in the Super-Kamiokande
atmospheric neutrino experiment
and in solar neutrino
experiments
imply that the law of conservation of family lepton numbers $L_{e}$, $L_{\mu}$
and $L_{\tau}$ is strongly violated. 
We consider the
states of flavor neutrinos  $\nu_{e}$, $\nu_{\mu}$ and
$\nu_{\tau}$
and we discuss the evolution
of these states in space and time
in the case of non-conservation of
family lepton numbers due to the mixing
of light neutrinos.
We discuss and compare different flavor neutrino discovery
experiments.
We stress that
experiments on the search for
$\nu_\mu\to\nu_\tau$ and $\nu_e\to\nu_\tau$
oscillations
demonstrated that the
flavor neutrino $\nu_{\tau}$ is a new type of neutrino, different from
$\nu_{e}$ and $\nu_{\mu}$.
In the case of neutrino mixing, the lepton number (only one)
is connected with the nature of massive neutrinos.
Such conserved lepton number exist if massive neutrinos are Dirac
particles.
We review possibilities to check in future experiments whether
the conserved lepton number exists.
\end{abstract}
\end{titlepage}

\section{Introduction}
\label{Introduction}

The strong evidence in favor of neutrino masses and mixing that was
obtained recently in the Super-Kamiokande atmospheric neutrino
experiment
\cite{SK-atm}
opened a new epoch in neutrino physics.
Evidence in favor of neutrino mixing was also obtained in all solar
neutrino experiments:
Homestake \cite{Homestake-98},
Kamiokande \cite{Kamiokande-sun-96},
GALLEX \cite{GALLEX-99},
SAGE \cite{SAGE-99},
Super-Kamiokande \cite{SK-sun},
GNO \cite{GNO}.
Indications in favor
of  $\nu_{\mu} \to \nu_{e}$ oscillations were found in the 
accelerator LSND experiment \cite{LSND}.

All these data can be explained
in terms of neutrino oscillations,
if the masses of neutrinos are different
from zero and the fields of massive neutrinos enter in the standard 
CC and NC interaction Lagrangian
\begin{align}
\null & \null
\mathcal{L}_{I}^{\mathrm{CC}}
=
- \frac{g}{2\sqrt{2}}
\,
j_{\alpha}^{\mathrm{CC}}
W^{\alpha}
+
\text{h.c.}
\,,
\quad
\null && \null
j_{\alpha}^{\mathrm{CC}}
=
2
\sum_{\ell=e,\mu,\tau}
\overline{\nu_{\ell L}} \, \gamma_{\alpha} \, \ell_L
+
\ldots
\label{001}
\\
\null & \null
\mathcal{L}_{I}^{\mathrm{NC}}
=
- \frac{g}{2\cos\theta_{W}}
\,
j_{\alpha}^{\mathrm{NC}}
Z^{\alpha}
+
\text{h.c.}
\,,
\quad
\null && \null
j_{\alpha}^{\mathrm{NC}}
=
\sum_{\ell=e,\mu,\tau}
\overline{\nu_{\ell L}} \, \gamma_{\alpha} \, \nu_{\ell L}
+
\ldots
\label{002}
\end{align}
in the mixed form\footnote{
In order to describe all existing neutrino oscillation data,
including LSND data, it is necessary to assume that there are
transitions of flavor neutrinos
$\nu_{\ell}$ into sterile states
(see, for example, \cite{BGG-review-98}).
We will not consider here this possibility.
}
\begin{equation}
\nu_{\ell L}
=
\sum_{i=1}^{3} U_{\ell i} \, \nu_{iL}
\,.
\label{003}
\end{equation}
Here $\nu_{i}$ is the field of the neutrino with mass $m_{i}$ and
$U$ is the mixing matrix.  

In the framework of theories with massless neutrinos it was customary to 
introduce the family lepton numbers
$L_{e}$, $L_{\mu}$ and $L_{\tau}$, correspondingly, for the pairs       
($\nu_{e},e^{-}$),
($\nu_{\mu},\mu^{-}$)
and
($\nu_{\tau},\tau^{-}$).
The observation of neutrino oscillations clearly demonstrates that
family lepton numbers are not conserved. 

In Section \ref{Flavor neutrino states}
we consider the states of flavor neutrinos and the
evolution of these states in vacuum.
In Section \ref{Flavor neutrino discovery experiments}
we discuss and compare different flavor neutrino
discovery experiments.
In Section \ref{Lepton number and neutrino mixing}
we review the
concept of lepton number in the framework of
the theory of neutrino masses and mixing.

\section{Flavor neutrino states}
\label{Flavor neutrino states}

In this Section we consider,
in the framework of neutrino mixing:
\begin{enumerate} 
\item
Decays in which family lepton numbers are not conserved  
(like
$\mu^{+} \to e^{+} + \gamma$,
$\mu^{+} \to e^{+} + e^{-} + e^{+}$
and others).
\item
Flavor neutrino states.
\item
Transitions between flavor neutrinos in vacuum (neutrino
oscillations).
\end{enumerate}

\begin{figure}[p!]
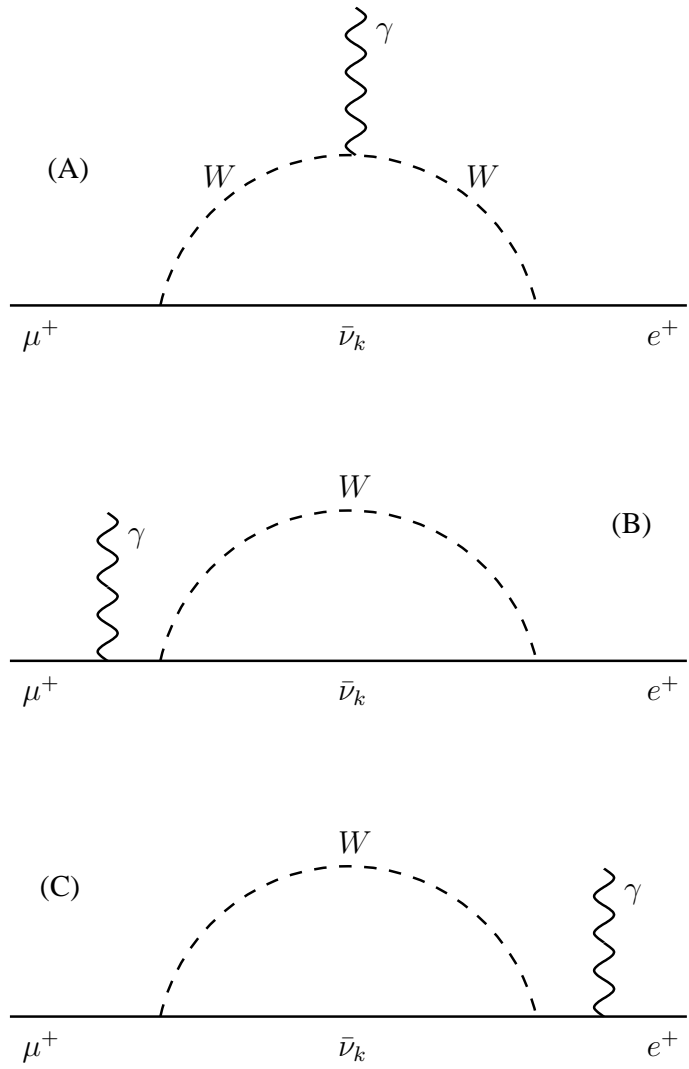

\begin{center}
\begin{tabular}{c}
\input{meg-a.pstex_t}
\\
\null\vspace{1cm}\null
\\
\input{meg-b.pstex_t}
\\
\null\vspace{1cm}\null
\\
\input{meg-c.pstex_t}
\end{tabular}
\end{center}
\caption{ \label{meg}
Diagrams contributing to the decay
$\mu^{+} \to e^{+} + \gamma$
at lowest order.
}
\end{figure}

If there is neutrino mixing
(Eq.~(\ref{003})),
processes 
of transitions between leptons of different families,
as
$\mu^{+} \to e^{+} + \gamma$,
$\mu^{+} \to e^{+} + e^{-} + e^{+}$ 
and others,
become possible.
Let us consider, for example,
the decay
$\mu^{+} \to e^{+} + \gamma$
(see the diagrams in Fig.\ref{meg}).
In the simplest case of
mixing of two neutrinos with masses $m_{1}$
and $m_{2}$ the ratio $R$ of the probability of
$\mu^{+} \to e^{+} + \gamma$ decay
to the probability of the decay
$\mu^{+} \to e^{+} + \nu_{e} + \bar\nu_{\mu}$
is given by \cite{Petcov:1977ff,Cheng:1980tp,Bilenky-Petcov-RMP-87}
\begin{equation}
R
=
\frac{3\alpha}{128\pi}
\left( \frac{\Delta{m}^2}{m_{W}^2} \right)^2
\sin^2 2\vartheta
\,,
\label{004}
\end{equation}
where $m_{W}$ is the mass of the $W$ boson,
$\Delta{m}^{2} \equiv m_{2}^{2} -  m_{1}^{2}$
and $\vartheta$ is the neutrino mixing angle.

The value of $\Delta{m}^{2}/m_{W}^{2}$
in Eq.~(\ref{004}) is
\begin{equation}
\frac{\Delta{m}^{2}}{m_{W}^{2}}
\simeq
1.5 \times 10^{-22} \, \frac{\Delta{m}^{2}}{\mathrm{eV}^2}
\,.
\label{005}
\end{equation}
For $\Delta{m}^{2} \lesssim 1 \, \mathrm{eV}^{2}$, for the ratio $R$ we have
\begin{equation}
R \lesssim 1.2 \times 10^{-48}
\,.
\label{006}
\end{equation}
Thus, though the processes 
$\mu^{+} \to e^{+} + \gamma$, 
$\mu^{+} \to e^{+} + e^{-} + e^{+}$ and others
are in principle allowed 
in the case of neutrino mixing,
it is practically
impossible to observe them\footnote{
If the violation of the law of conservation
of lepton numbers is due to other mechanisms,
as supersymmetry, the probability of the decay
$\mu^{+} \to e^{+} + \gamma$ 
and other similar processes can be much larger than in the case 
of neutrino mixing
(see \cite{Barbieri:1994pv,Barbieri:1995tw,%
Arkani-Hamed:1996fs,Hisano:1997qq}).
}.

The strong suppression of the probability of processes like
$\mu^{+} \to e^{+} + \gamma$
is due to the fact that the
coefficient in Eq.~(\ref{005}) is very small.
As we will see later,
in the case of neutrino oscillations
the corresponding
coefficient, $L / E$
($E$ is the neutrino energy and $L$ is the distance
between neutrino source and detector),
can be many orders of magnitude larger.
This is the main reason why effects of
violation of the law of conservation of lepton numbers can be revealed in
neutrino oscillation experiments\footnote{
In the case of neutrinoless double-$\beta$ decay of nuclei
$\mathcal{N}(A,Z) \to \mathcal{N}(A,Z+2) + e^{-} + e^{-}$
the suppression
of the decay probability
is less strong than
in the decay
$\mu^{+} \to e^{+} + \gamma$
and similar processes.
We consider neutrinoless double-$\beta$ decay
in Section~\ref{Lepton number and neutrino mixing}.
}.

Let us consider now,
in the framework of neutrino mixing,
the state
$|\nu_\ell\rangle$
of a flavor neutrino produced in the CC
weak decay process
\begin{equation}
A \to B + \ell^+ + \nu_\ell
\,.
\label{0100}
\end{equation}
For example,
in the case of neutrinos produced in nuclear $\beta^+$
decay,
$A = \mathcal{N}(A,Z)$,
$B = \mathcal{N}(A,Z-1)$,
$\ell=e$.

If there is neutrino mixing,
the flavor neutrino state
$|\nu_\ell\rangle$
is a superposition of states of massive neutrinos $\nu_i$:
\begin{equation}
|\nu_{\ell}\rangle
=
\sum_i
|\nu_i\rangle
\langle\ell^{+},\nu_i,B|S|A\rangle
\,,
\label{007}
\end{equation}
where
$\langle\ell^{+},\nu_i,B|S|A\rangle$
is the relevant element of the $S$ matrix.

From the data of laboratory experiments
and
astrophysical observations we know that
neutrino masses are very small
(see \cite{PDG}),
\begin{equation}
m_i \lesssim \text{few eV}
\,.
\label{010}
\end{equation}
Since in order to be detected
in present-day experiments
neutrinos must have energy
$E \gtrsim 0.1 \, \mathrm{MeV}$,
we have
$m_{i}^{2}/E^{2} \ll 1$.
Thus,
the kinematical dependence of the matrix element
$\langle\ell^{+},\nu_i,B|S|A\rangle$
on neutrino masses can be neglected with very good approximation,
leading to
\begin{equation}
\langle\ell^{+},\nu_i,B|S|A\rangle
\simeq
U_{\ell i}^*
\left.
\langle\ell^{+},\nu_\ell,B|S|A\rangle
\right|_{m_i=0}
\,.
\label{01101}
\end{equation}
Here
$
\left.
\langle\ell^{+},\nu_\ell,B|S|A\rangle
\right|_{m_i=0}
$
is the matrix element evaluated in the Standard Model,
with zero neutrino masses.
From Eqs.~(\ref{007}) and (\ref{01101}),
the normalized state describing the flavor neutrino
produced in the decay process (\ref{0100}) is
\begin{equation}
|\nu_{\ell}\rangle
=
\sum_{i} U_{\ell i}^* \, |\nu_i\rangle
\,.
\label{014}
\end{equation}

Neutrinos $\nu_{i}$ with mass $m_{i}$ are produced in
standard weak
decays in states with left-handed as well as 
right-handed helicities.
However, in the Standard Model the probability to produce
a neutrino $\nu_{i}$ in a state with
right-handed helicity is
negligibly small because it is proportional to $m_{i}^{2}/E^{2}$.
Thus, $\nu_{i}$'s and, consequently, flavor neutrinos
$\nu_{\ell}$ are produced in standard weak interaction
processes
in almost pure left-handed states.

The state of antineutrino
$|\bar{\nu}_{\ell}\rangle$,
the particle that is produced in a CC
weak process together with a $\ell^{-}$,
is given by
\begin{equation}
|\bar{\nu}_{\ell}\rangle
=
\sum_i U_{\ell i} |\bar\nu_i\rangle
\,.
\label{015}
\end{equation}
The state vector $|\bar\nu_{i}\rangle$ describes
antineutrinos with right-handed helicity in the case of Dirac $\nu_{i}$
or
neutrinos with right-handed helicity in the case of Majorana $\nu_{i}$.
Hence,
the state vector $|\bar{\nu}_{\ell}\rangle$
describes neutrinos with right-handed helicity.

In the general case of CP violation in the lepton sector,
there are phases in the neutrino
mixing matrix $U$.  
Therefore,
the states $|\nu_{\ell}\rangle$ and  $|\bar{\nu}_{\ell}\rangle$ 
differ not only by helicity, but also by the sign of the
CP-violating phases.
The violation of CP in the lepton sector can be
revealed through the investigation of neutrino oscillations
(see, for example, \cite{Bilenky-Petcov-RMP-87,BGG-review-98}).

In oscillation experiments neutrinos
are detected at some distance from
the source.
Neutrinos produced as flavor neutrinos $\nu_{\ell}$
are described at the source by the state (\ref{014}).
Taking into account the evolution
in space and time,
the neutrino beam
at the distance $\vet{x}$ from the source
and
at the time $t$ after production
is described by the state
\begin{equation}
\sum_{i}
e^{i ( \vet{p}_i \cdot \vet{x} - E_i t )}
\,
U_{\ell i}^*
\,
|\nu_i\rangle
\,,
\label{016}
\end{equation}
where
$\vet{p}_i$ is the three-momentum of the massive neutrino $\nu_i$
and
$E_i = \sqrt{|\vet{p}_i|^2 + m_i^2 }$
is its energy.

Let us consider
an experiment in which neutrinos
described by the state (\ref{016})
are detected through the observation of
the CC process
\begin{equation}
\nu_{\ell'} + N \to \ell' + X
\,.
\label{0172}
\end{equation}
Here $N$ is a target nucleon,
$\ell'$ is the final lepton and
$X$ represents final hadrons.
The amplitude of the process (\ref{0172}) is proportional to
\begin{equation}
\sum_{i}
\langle \ell' , X | S | N , \nu_i \rangle
\,
e^{i ( \vet{p}_i \cdot \vet{x} - E_i t )}
\,
U_{\ell i}^*
\,.
\label{018}
\end{equation}
As in the case of neutrino production,
for ultra-relativistic neutrinos
the contribution of neutrino masses
to the matrix element in Eq.~(\ref{018})
can be neglected with very good approximation:
\begin{equation}
\langle \ell' , X | S | N , \nu_i \rangle
\simeq
U_{\ell' i}
\left.
\langle \ell' , X | S | N , \nu_{\ell'} \rangle
\right|_{m_i=0}
\,,
\label{019}
\end{equation}
where
$
\left.
\langle \ell' , X | S | N , \nu_{\ell'} \rangle
\right|_{m_i=0}
$
is the matrix element evaluated in the Standard Model,
with massless neutrinos.
Therefore, the probability amplitude to observe a
flavor neutrino
$\nu_{\ell'}$
at the distance $\vet{x}$ from the source
and
at the time $t$ after production of a flavor neutrino $\nu_{\ell}$
is given by
\begin{equation}
\mathcal{A}_{\nu_{\ell}\to\nu_{\ell'}}(\vet{x},t)
=
\sum_{i}
U_{\ell' i}
\,
e^{i ( \vet{p}_i \cdot \vet{x} - E_i t )}
\,
U_{\ell i}^*
\,.
\label{021}
\end{equation}
From this expression it is clear that
transitions between
different flavor neutrinos can take place only
if the following two conditions are satisfied:
\begin{enumerate}
\item 
The matrix $U$ is non-diagonal.
\item
The phase factors $e^{i ( \vet{p}_i \cdot \vet{x} - E_i t )}$
for different massive
neutrinos $\nu_{i}$ are different.
\end{enumerate} 
If neutrinos are massless,
$\vet{p}_i = \vet{p}$ and $E_i=E$,
leading to the transition probability
\begin{equation}
P_{\nu_{\ell}\to\nu_{\ell'}}(\vet{x},t)
=
|\mathcal{A}_{\nu_{\ell}\to\nu_{\ell'}}(\vet{x},t)|^2
=
\delta_{\ell\ell'}
\,.
\label{021001}
\end{equation}
In general,
using the unitarity of the neutrino mixing matrix,
we have
\begin{equation}
0
\leq
P_{\nu_{\ell}\to\nu_{\ell'}}(\vet{x},t)
\leq
\sum_{i} |U_{\ell' i}|^2
\,
\sum_{k} |U_{\ell k}|^2
=
1
\,.
\label{021002}
\end{equation}

Let us enumerate the neutrino masses in such a way that
\begin{equation}
m_{1} <  m_{2} < m_{3}
\,.
\label{022}
\end{equation}
Taking into account the unitarity of the mixing matrix $U$,
the amplitude (\ref{021}) of
$\nu_{\ell} \to \nu_{\ell'}$
transitions
can be written in the form
\begin{equation}
\mathcal{A}_{\nu_{\ell}\to\nu_{\ell'}}(x)
=
e^{i ( \vet{p}_1 \cdot \vet{x} - E_1 t )}
\left\{
\delta_{\ell \ell'}
+
\sum_{i}
U_{\ell i}^*
U_{\ell' i}
\left[
e^{i ( \vet{p}_i - \vet{p}_1 ) \cdot \vet{x} - i ( E_i - E_1 ) t )}
-
1
\right]
\right\}
\,.
\label{023}
\end{equation}
It is obvious that the common phase
$e^{i ( \vet{p}_1 \cdot \vet{x} - E_1 t )}$
does not enter into the expression
for the transition probability.
Taking into account the fact that the neutrino is detected only
if its three-momentum is aligned along $\vet{x}$
(\textit{i.e.}
$\vet{p} = |\vet{p}| \vet{x} / |\vet{x}|$),
the phase difference
$\phi_{i1} = ( \vet{p}_i - \vet{p}_1 ) \cdot \vet{x} - ( E_i - E_1 ) t$
in Eq.~(\ref{023})
can be written as
\begin{equation}
\phi_{i1}
=
( p_i - p_1 ) x - ( E_i - E_1 ) t
\,,
\label{02401}
\end{equation}
where $x \equiv |\vet{x}|$
and
$p_i \equiv |\vet{p}_i|$.
Furthermore,
we have
\begin{equation}
\phi_{i1}
=
\frac{p_i^2 - p_1^2}{p_i + p_1} \, x
- (E_i - E_1) t
=
(\phi_{i1})_{\mathrm{st}}
- (E_i - E_1)
\left( t - \frac{E_i + E_1}{p_i + p_1} \, x \right)
\,,
\label{025}
\end{equation}
where
\begin{equation}
(\phi_{i1})_{\mathrm{st}}
=
- \frac{\Delta{m}^2_{i1} x}{p_i + p_1}
\simeq
- \frac{\Delta{m}^2_{i1} x}{2E}
,
\label{02601}
\end{equation}
with
$\Delta{m}^2_{ij} \equiv m_i^2 - m_j^2$,
is the standard expression for the 
phase difference
(see, for example, \cite{Bilenky-Pontecorvo-PR-78,%
Bilenky-Petcov-RMP-87,%
CWKim-book,%
BGG-review-98}).
Here
$E$ is the neutrino energy given
by the kinematics of the production process
neglecting neutrino masses.
The second term on the right-hand side of Eq.~(\ref{025})
is much smaller than $(\phi_{i1})_{\mathrm{st}}$.
Indeed,
the kinematics of the production process
implies that
\begin{equation}
E_i - E_1
\sim
\frac{\Delta{m}^2_{i1}}{E}
\,.
\label{026}
\end{equation}
Since the
velocity of the neutrino signal
is equal to the velocity of light
minus a correction of the order
$m_i^2/E^2$,
we have
\begin{equation}
\left( t - \frac{E_i + E_1}{p_i + p_1} \, x \right)
\sim
\frac{m_i^2}{E^2} \, x
\,.
\label{028}
\end{equation}
Therefore,
the second term on the right-hand side of Eq.~(\ref{025})
is of the order
\begin{equation}
\left|(\phi_{i1})_{\mathrm{st}}\right|
\,
\frac{m_i^2}{E^2}
\ll
\left|(\phi_{i1})_{\mathrm{st}}\right|
\,,
\label{029}
\end{equation}
and
can be neglected\footnote{
Let us stress that we did not assume the
equality of momenta or equality of energies of massive neutrinos $\nu_{i}$.
Such assumptions are often discussed in literature
(see \cite{Giunti-Kim-QMNO-00,Okun-00,Grimus-Mohanty-Stockinger-strength-99}
and references therein).
The wave packet treatment
of neutrino transitions gives the same result
\cite{Giunti-Kim-Lee-Whendo-91,Giunti-Kim-Coherence-98}.
}.

The probability of $\nu_{\ell} \to \nu_{\ell'}$
transitions is given by
\begin{equation}
P_{\nu_{\ell}\to\nu_{\ell'}}(L)
=
\left|
\delta_{\ell \ell'}
+
\sum_{i}
U_{\ell i}^*
U_{\ell' i}
\left(
e^{-i \frac{\Delta{m}^2_{i1}L}{2E}}
-
1
\right)
\right|^2
\,,
\label{02301}
\end{equation}
where
$L=x$
is the distance between neutrino source and detector.

The transition probability depends on the quantity $L/E$ that
is determined by the experimental conditions.
If the quantity  $L/E$ is so small that for all 
$\Delta{m}^{2}_{i1}$
\begin{equation}
\frac{\Delta{m}^2_{i1}L}{2E}
\ll
1
\,,
\label{024}
\end{equation}
then
$P(\nu_{\ell} \to \nu_{\ell'}) \simeq \delta _{\ell\ell'}$
and violation of the law of conservation of family lepton numbers
cannot be observed.
A violation of this law
can be observed only if
the quantity  $L/E$ is large enough so that for at least one
neutrino mass-squared difference,
say $\Delta{m}^2$,
\begin{equation}
\frac{\Delta{m}^2 L}{2E}
\gtrsim
1
\,.
\label{02402}
\end{equation}
This condition can be rewritten as
\begin{equation}
2.5
\
\frac{L (\mathrm{m})}{E (\mathrm{MeV})}
\
\Delta{m}^2 (\mathrm{eV}^2)
\gtrsim
1
\,.
\label{02502}
\end{equation}
where   
$L (\mathrm{m})$ is the source-detector distance in meters,
$E (\mathrm{MeV})$ is neutrino energy in MeV,
and
$\Delta{m}^2 (\mathrm{eV}^2)$
is the neutrino mass-squared difference in $\mathrm{eV}^2$.
In short-baseline neutrino oscillation experiments
$L/E \sim 10^{-2}-10^2$,
in long-baseline neutrino oscillation experiments
$L/E \sim 10^{2}-10^{3}$,
in atmospheric neutrino experiments
$L/E \sim 10^{2}-10^{4}$,
and
in solar neutrino experiments  
$L/E \sim 10^{10}-10^{11}$,
leading to a sensitivity to
$\Delta{m}^2 \sim 10^{-2}-10^2 \, \mathrm{eV}^2$,
$10^{-3}-10^{-2} \, \mathrm{eV}^2$,
$10^{-4}-10^{-2} \, \mathrm{eV}^2$,
$10^{-11}-10^{-10} \, \mathrm{eV}^2$,
respectively.

In conclusion of this section,
let us stress that
in the case of
\emph{mixing of neutrinos with small masses},
flavor neutrinos and antineutrinos
are not quanta of the
$\nu_{e}$,
$\nu_{\mu}$
and
$\nu_{\tau}$
fields
\cite{Giunti-Kim-Lee-Remarks-92}.
In other words,
$\nu_{e}$,
$\nu_{\mu}$
and
$\nu_{\tau}$
are not fields of particles\footnote{
There is no difference of principle between neutrino mixing and quark
mixing. It is obvious that, for example, in the quark case
\begin{equation*}
d'_L
=
\sum_{q=d,s,b} V_{uq} \, q_L
\end{equation*}
is not a particle field but
the combination of the left-handed components
of the $d$, $s$ and $b$ fields.
}.

The neutrinos $\nu_{e}$, $\nu_{\mu}$, $\nu_{\tau}$ 
(antineutrinos $\bar\nu_{e}$, $\bar\nu_{\mu}$, $\bar\nu_{\tau}$),
are produced
in CC weak decays together with, correspondingly,
$e^{+}$,
$\mu^{+}$,
$\tau^{+}$ 
($ e^{-}$,  $\mu^{-}$,  $\tau^{-}$),
and
produce, correspondingly,
$ e^{-}$,  $\mu^{-}$  $\tau^{-}$
($e^{+}$,  $\mu^{+}$, $\tau^{+}$)
in CC processes of the
interaction with nucleon etc.
These neutrinos carry the flavor of corresponding leptons and 
their appropriate names are \emph{flavor neutrinos}\footnote{
Sometimes states of flavor neutrinos are called eigenstates of weak
interactions. We do not think that this name reflect the real content
of the notion of a flavor neutrino state.
}.

The states of flavor neutrinos (antineutrinos) are 
the superpositions of
states of neutrinos with definite masses and negative helicity 
(positive helicity).
Thus, \emph{flavor neutrinos do not have definite mass}.

The investigation of neutrino oscillations is the most sensitive method
to reveal the violation of the law of conservation of family lepton numbers
(see \cite{Bilenky-Pontecorvo-PR-78,%
Bilenky-Petcov-RMP-87,%
CWKim-book,%
BGG-review-98}).

\section{Flavor neutrino discovery experiments}
\label{Flavor neutrino discovery experiments}

In this section
we discuss and compare different flavor neutrino
discovery experiments.

As it is well known,
the electron neutrino $\nu_{e}$ was
discovered  by C.L. Cowan and F. Reines in the fifties
\cite{Reines:1953pu,Cowan:1956gi,Reines:1959nc}.
In 1962, in the Brookhaven experiment
of Lederman, Schwartz, Steinberger \textit{et al.}
\cite{Danby:1962nd}
the second flavor neutrino $\nu_{\mu}$ was discovered.
In 2000
the tau neutrino $\nu_{\tau}$
has been directly detected
in the DONUT experiment \cite{DONUT}.

In the Cowan and Reines experiment,
electron (anti)neutrinos have been detected through
the observation of the process
\begin{equation}
\bar{\nu}_e + p \to e^{+} + n
\,,
\label{03101}
\end{equation}
with antineutrinos from the powerful Savannah River reactor. 

In a reactor $\bar{\nu}_e$'s are produced in a chain of
$\beta$-decays
of radioactive neutron-rich nuclei, products
of the fission of $^{235}$U and  $^{238}$U. 
The energy of reactor antineutrinos
is less than about 10 MeV.
About $2 \times 10^{14}$ 
$\bar{\nu}_e$'s are emitted per second per KW.
The power of the Savannah River
reactor was $\simeq 2300 \, \mathrm{MW}$ (th).
Thus, about  $2.3 \times 10^{20}$
$\bar\nu_e$'s per second were emitted by the reactor.
The flux of $\bar{\nu}_e$'s in the Cowan and Reines experiment was
$\simeq 10^{13} \, \mathrm{cm}^{-2} \, \mathrm{s}^{-1}$.

As it is well known,
the hypothesis of the existence of neutrino was
put forward in 1930
by W. Pauli in order to solve the problem of continuous 
$\beta$-spectra and the problem of the spin and statistics
of some nuclei (like $^{14}N$).
In 1933 E. Fermi assumed that
an electron and an antineutrino are produced in the
process
\begin{equation}
n \to p + e^{-} + \bar{\nu}_e
\,,
\label{032}
\end{equation}
and proposed the first Hamiltonian of $\beta$-decay.

It is a \emph{direct consequence of quantum field theory}
that an
$\bar{\nu}_e$ that is produced in $\beta$-decay together
with an electron
\emph{must} produce a positron in the process (\ref{03101}).
Moreover, if the interaction responsible for the decay of the neutron is
known, one can connect the cross
section of the process (\ref{03101}) at the small reactor energies
with the lifetime of the neutron.
Neglecting small corrections due to neutron recoil,
the total cross
section of the process (\ref{03101}) is given by
(see, for example, \cite{Konopinski-66})
\begin{equation}
\sigma(\bar{\nu}_e p \to e^{+} n)
=
\frac{2 \pi^2}{m_e^5 f \tau_n} \, p_e \, E_e
\,.
\label{033}
\end{equation}
Here 
$E_{e} = E - (m_{n} - m_{p})$
is the energy of the positron
($E$ is the antineutrino energy),
$\tau_{n} = 886.7 \pm 1.9 \, \mathrm{s}$ is the lifetime of the neutron,
$f= 1.686$ is the neutron 
statistical factor that includes Coulomb interactions of the final proton
and electron,
$m_n$, $m_p$ and $m_e$ are masses of
the neutron, proton and electron, respectively.

In the Cowan and Reines experiment 
positrons and neutrons produced in the process
(\ref{03101})
were detected
and
for the first time the corresponding very
small neutrino cross section
$\sigma \sim 10^{-43} \, \mathrm{cm}^{2}$
was measured.
This became possible because of the existence of an
intensive source
of antineutrinos (reactor) and because of the invention of
large scintillator counters.

The total cross section of the process (\ref{03101}) measured in the
Cowan and Reines experiment,
\begin{equation}
\sigma(\bar{\nu}_e p \to e^{+} n)_{\mathrm{exp}}
=
(11 \pm 4) \times 10^{-44} \, \mathrm{cm}^{2}
\,,
\label{034}
\end{equation}
was in an agreement with the expected cross section
\begin{equation}
\sigma(\bar{\nu}_e p \to e^{+} n)_{\mathrm{th}}
=
9.5 \times 10^{-44} \, \mathrm{cm}^{2}
\,.
\label{035}
\end{equation}
The Cowan and Reines experiment
was a crucial confirmation of the
Pauli-Fermi hypothesis of existence of the neutrino.
This experiment also confirmed the correctness of the field-theoretical
relation (\ref{033})
between the lifetime the neutron and the cross section of the
cross-symmetrical process (\ref{03101}).
However,
since the energy of antineutrinos from a reactor is not
enough to produce muons, the Cowan and Reines experiment
could not reveal the existence of other flavor neutrinos, besides
$\nu_{e}$.

The next flavor neutrino discovery experiment was the 1962 Brookhaven
experiment
of L.M. Lederman, M. Schwartz, J. Steinberger \textit{et al.}
\cite{Danby:1962nd}.

At that time there were some indications that the muon neutrino
(the neutrino that is produced in $\mu$-capture,
$\mu$-decay
and other weak processes in which the muon participates)
and the electron neutrino are different particles.
These indications were based
on the comparison of the results of calculations of the
probability
of $\mu \to e + \gamma$ decay with
the experimental upper bound for the
probability of this decay.

If $\nu_{\mu}$ and $\nu_{e}$ are the same particle,
the decay $\mu \to e + \gamma$ is allowed.
The probability of this decay was calculated in
Ref.~\cite{Feinberg-58}
in the framework of a nonrenormalizable theory with
intermediate $W$ boson
(diagrams are similar to the diagrams in Fig.~\ref{meg}),
assuming  
that the $W$-boson has a normal magnetic
moment and that the cut-off mass
is equal to the mass of the  $W$.
The resulting value of
the ratio $R$ of the probability of the
decay $\mu \to e + \gamma$
and
the total probability of muon decay
was $R = \alpha/(24\pi) \simeq 10^{-4}$
\cite{Feinberg-58}.

On the other hand,
the decay $\mu \to e + \gamma$ was not observed experimentally.
At the time of the Brookhaven experiment,
the upper bound
was $R \lesssim 10^{-8}$
\cite{Bartlett-62,Frankel-62}\footnote{
Now the upper bound
is $R \leq 1.2\times 10^{-11}$
\cite{PDG}. 
}.

In spite of the indication of the existence of a muon neutrino
given by the non-observation of
$\mu \to e + \gamma$
decays,
it was extremely important
to check whether $\nu_{\mu}$ and $\nu_{e}$ are the same
or different particles in a \emph{direct neutrino experiment}.
The  Brookhaven experiment was the ideal experiment
for this aim\footnote{
The experiment was proposed by B. Pontecorvo in 1959
\cite{Pontecorvo-neutrinos-59}.
}.

In this experiment the neutrino beam
was produced in the decays of pions
with a small admixture of neutrinos from the decays of kaons and muons.
The dominant decay mode of the $\pi^{+}$ meson is
\begin{equation}
\pi^{+} \to \mu^{+} + \nu_{\mu}
\,.
\label{041}
\end{equation}
According to the universal $V-A$ theory of weak interactions of
Feynman and Gell-Mann \cite{Feynman-Gell-Mann-58}
and
Marshak and Sudarshan \cite{Sudarshan-Marshak-58},
the ratio of the probability of the decay
\begin{equation}
\pi^{+} \to e^{+} + \nu_{e}
\label{042}
\end{equation}
and the probability of the decay in Eq.~(\ref{041})
is about\footnote{
This prediction of the V-A
theory was beautifully confirmed in a CERN experiment in 1958
\cite{Fazzini:1958ii}.
}
$1.2 \times 10^{-4}$.
Hence, the neutrino beam in the Brookhaven experiment was
practically a pure beam of muon neutrinos.

The pion beam in the experiment was produced by 15 GeV
protons striking a berillium target.
Neutrinos from the decays of pions had a
spectrum of energies $\lesssim 1 \, \mathrm{GeV}$.
Neutrino interactions were observed in a 10 ton
aluminum spark chamber.

According to field theory,
a muon neutrinos produced in the decay
(\ref{041})
together with a $\mu^{+}$ \emph{must} produce 
$\mu^{-}$ in the process
\begin{equation}
\nu_{\mu} + N \to \mu^{-} + X
\,.
\label{043}
\end{equation}
In order to investigate if
$\nu_{\mu}$ and $\nu_{e}$
are the same or different particles,
one needs to check whether
$\nu_{\mu}$'s can produce also electrons in
the process
\begin{equation}
\nu_{\mu} + N \to e^{-} + X
\,.
\label{044}
\end{equation}
If electron and muon neutrinos are the same particles,
according to the universal $V-A$ theory one must expect to
observe in the detector approximately
an equal number of electrons and muons.

In the Brookhaven experiment 34 single muon events have been observed,
with an expected background from cosmic rays of 5 events.
The measured cross
section was in agreement with the $V-A$ theory.
Six shower events were observed,
with a distribution of sparks totally different from that
expected for electrons.
If $\nu_{\mu}$ and  $\nu_{e}$ are the same particles,
29 electron events with energy more than 400 MeV should have been
observed in the experiment\footnote{
In 1963 in CERN, with the invention of the magnetic horn, the intensity
and purity of neutrino beams was greatly improved.
The Brookhaven result was confirmed with good accuracy
in a large 45 tons
spark-chamber experiment
\cite{Bienlein-64}
and in a large bubble chamber experiment
\cite{Block-64}.
}.

Summarizing,
the Brookhaven experiment proved that muon neutrinos,
produced together with muons, cannot produce electrons in 
the process (\ref{044}).
Therefore, it was proved that
\emph{$\nu_{e}$ and $\nu_{\mu}$ are different flavor neutrinos}.
The Brookhaven experiment also proved for the first time that
accelerator $\nu_\mu$'s produced in the process (\ref{041})
can be detected.
      
Let us notice that the results of the Brookhaven experiment
and all other data existing at that time were interpreted in terms
of two
conserved family lepton numbers $L_{e}$ and $L_{\mu}$ that allowed to
distinguish $( \nu_{e}, e^{-} )$ and $( \nu_{\mu}, \mu^{-} )$ pairs
and to forbid processes of type (\ref{044}).
We know now that in the framework of neutrino mixing family lepton
numbers are not conserved and muon neutrinos
at some distance can transform into electron neutrinos and produce
electrons (as in the case of LSND experiment \cite{LSND}).
From the point of view of
neutrino mixing, flavor neutrino discovery experiments require
relatively small distances between neutrino sources and detector and
relatively large energies,
in order to satisfy the condition (\ref{024}).

In 1975 the third lepton, $\tau$, was discovered by M. Perl \textit{et al.}
\cite{Perl:1975bf,Perl:1980pb}.
After this discovery many decay modes of $\tau$ have been investigated:
\begin{align}
\null & \null
\tau^- \to \mu^- + \bar\nu_\mu + \nu_\tau
\,,
\null && \null
\tau^- \to \pi^- + \nu_\tau
\,,
\nonumber
\\
\null & \null
\tau^- \to e^- + \bar\nu_e + \nu_\tau
\,,
\null && \null
\tau^- \to \pi^- + \pi^0 + \nu_\tau
\,,
\nonumber
\\
\null & \null
\null && \null
\tau^- \to K^- + \nu_\tau
\,,
\label{045}
\end{align}
and others.

All experimental data on $\tau$
decays are in good agreement with the Standard Model
\cite{Pich:2000rj,Perl:1998ht,Stahl:2000aq}.
It is a general a consequence of field theory that
the neutrino $\nu_\tau$ produced in $\tau$ decays
as those in Eq.~(\ref{045})
can produce  $\tau^{-}$'s in process as
\begin{equation}
\nu_{\tau} + N \to \tau^{-} + X
\label{046}
\end{equation}
and others.
Moreover,
$e-\mu-\tau$ universality of weak interactions
allows to predict the cross section of the
process (\ref{046}).

Therefore,
the investigation of $\tau$ decays of the type (\ref{045})
and subsequent charged-current processes as the one in Eq.~(\ref{046})
do not allow to check if
$\nu_{\tau}$ is a new type of neutrino, different from $\nu_{\mu}$
and $\nu_{e}$.
However,
as in the case of $\nu_{\mu}$ and  $\nu_{e}$
this can be tested in a different type of neutrino experiment.

In order to prove that $\nu_\tau$
is a new type of neutrino,
it is necessary to prove either that
$\nu_{\tau}$'s
cannot produce electrons or muons,
or that $\nu_{\mu}$'s and  $\nu_{e}$'s
cannot produce $\tau$'s.

The Brookhaven experiment proved that muon neutrinos produce
muons and do not produce
electrons with the predicted cross section.
However,
another type of experiment
could prove that
$\nu_{\mu}$ and $\nu_{e}$
are different particles.
Imagine that it would be possible to create a pure beam of $\nu_{e}$'s
with energies well above of the threshold of $\mu^{-}$ production.
If in an experiment with such a beam it were shown that
$\nu_{e}$'s
produce electrons and \emph{do not produce muons}
with the predicted cross section
(under the assumption that $\nu_{\mu}$ and $\nu_{e}$ are the same particles)
it would be proven
that the flavor neutrinos $\nu_{\mu}$ and $\nu_{e}$ are different.

So far no experiment has proved that
$\nu_{\tau}$'s
cannot produce electrons or muons,
but several neutrino oscillation experiments
looking for
$\nu_\mu\to\nu_\tau$ and $\nu_e\to\nu_\tau$
transitions have
\emph{proved that $\nu_{\mu}$'s and  $\nu_{e}$'s
cannot produce $\tau$'s}.
These experiments are:
FNAL-E531
($\nu_\mu\to\nu_\tau$ and $\nu_e\to\nu_\tau$)
\cite{FNAL-E531},
CHARM II
($\nu_\mu\to\nu_\tau$)
\cite{CHARM-II-nu_mu_to_nu_tau-94},
CCFR
($\nu_e\to\nu_\tau$)
\cite{CCFR-nu_e_to_nu_tau-99},
CHORUS
($\nu_\mu\to\nu_\tau$ and $\nu_e\to\nu_\tau$)
\cite{CHORUS-01}
NOMAD
($\nu_\mu\to\nu_\tau$ and $\nu_e\to\nu_\tau$)
\cite{NOMAD-00}.

For example,
the neutrino beam in the recent
CHORUS and NOMAD experiments,
produced with the CERN SPS accelerator,
was predominantly composed of $\nu_{\mu}$'s,
with small
$\bar{\nu}_{\mu}$,
$\nu_{e}$
and
$\bar\nu_{e}$
components.
The percentage of $\nu_{e}$ was about 0.9\%
and the contamination
of $\nu_{\tau}$ in the beam is negligible 
($\simeq 5\times10^{-6}$).
The average
energies of $\nu_{\mu}$ and $\nu_{e}$ are 27 GeV
and
40 GeV, respectively. Notice that the threshold of production of
$\tau$'s
in the process (\ref{046}) is 3.5 GeV.

No event of $\tau$-lepton production have been observed in the
CHORUS and NOMAD experiments
at a distance of about 600 m from the source,
leading to the following upper bounds
for the probabilities of
$\nu_{\mu} \to \nu_{\tau}$ and $\nu_{e} \to \nu_{\tau}$ 
transitions:
\begin{align}
\null & \null
P_{\nu_{\mu} \to \nu_{\tau}} \leq 3.4 \times 10^{-4}
\,,
\null && \null
P_{\nu_{e} \to \nu_{\tau}} \leq 2.6 \times 10^{-2}
\null && \null
\protect\cite{CHORUS-01}
\,,
\nonumber
\\
\null & \null
P_{\nu_{\mu} \to \nu_{\tau}} \leq 2.1 \times 10^{-3}
\,,
\null && \null
P_{\nu_{e} \to \nu_{\tau}} \leq 2.6 \times 10^{-2}
\null && \null
\protect\cite{NOMAD-00}
\,.
\label{051}
\end{align}
These very stringent limits imply that
\emph{the flavor neutrinos 
$\nu_{\mu}$ and $\nu_{e}$ are different from $\nu_{\tau}$}.
If
$\nu_{\mu}$ and $\nu_{\tau}$ were the same particle,
about 5014 one-$\mu$ events
(events with
one reconstructed $\mu^{-}$ from the decay
$\tau^{-} \to \mu ^{-} + \bar\nu_{\mu} + \nu_{\tau}$)
would have been observed in the CHORUS experiment.
In reality no event of this type was observed.
If $\nu_{e}$ and $\nu_{\tau}$ were the same particle,
about 23 events with a highly energetic $e^{-}$ from
the decay
$\tau^{-} \to e ^{-} + \bar\nu_{e} + \nu_{\tau}$
would have been observed in the
NOMAD experiment.
No event of this type was observed above the expected background.

Let us notice that also
the experimental upper bounds for the
relative probabilities of the decays
$ \tau^{-} \to \mu^{-} + \gamma $ and $ \tau^{-} \to e^{-} + \gamma $,
\begin{equation}
R_{\mu} \leq 1.1 \times 10^{-6}
\,,
\qquad
R_{e} \leq 2.7 \times 10^{-6}
\protect\cite{PDG}
\,,
\label{052}
\end{equation}
imply that
$\nu_{\tau}$ is different from
$\nu_{e}$ and $\nu_{\mu}$.
This follows from
an argument that is similar to the one explained above for the decay
$\mu \to e + \gamma$ at the time of the Brookhaven
experiment,
based on the smallness of the upper limits (\ref{052})
with respect to
the value $R\simeq 10^{-4}$
expected if
$\nu_{\mu}$ and $\nu_{\tau}$ or $\nu_{e}$ and $\nu_{\tau}$ 
are the same particle \cite{Feinberg-58}.

The flavor neutrino $\nu_{\tau}$ was directly detected for the first
time in the DONUT experiment \cite{DONUT}. 
The DONUT experiment is a beam-dump experiment. 
Neutrinos in this experiment
were produced in the decays of short-lived charm particles.
The neutrino beam was composed mainly
of $\nu_{e}$'s and  $\nu_{\mu}$'s,
with about 5\% of $\nu_{\tau}$'s from the decay
\begin{equation}
D_{s} \to \tau + \nu_{\tau}
\,.
\label{053}
\end{equation}
Neutrinos are detected in the DONUT experiment in emulsions
at a distance of 36 m from the source.
The important signature of $\tau$ production is the kink from
$\tau$-decay.  
In a set of 203 neutrino interactions, four events with a kink,
which satisfy all requirements for the production and decay of $\tau$,
were found.
The estimated background is $0.34\pm 0.05$ events.

Up to now we considered only CC processes due to the intermediate
$W$-boson.
The investigation of NC processes due to the
intermediate $Z$-boson, as
\begin{equation}
\nu_{\mu} + N \to \nu_{\mu} + X
\,,
\quad
\nu_{\mu} + e \to \nu_{\mu} + e
\,,
\quad
\bar\nu_e + e \to \bar\nu_e + e
\,,
\label{054}
\end{equation}
and others,
have allowed to prove that
$\nu_{\mu}$ and $\nu_{e}$ interact with the $Z$-boson
in accordance with the Standard Model.

The four LEP experiments
(ALEPH, DELPHI, L3, OPAL)
determined with high accuracy that the number
$n_{\nu}$
of light flavor neutrinos
(mass $\lesssim 45 \, \mathrm{GeV}$)
produced in the decay of the $Z$-boson is three
(see \cite{PDG}):
\begin{equation}
n_{\nu} = 3.00 \pm 0.06
\,.
\label{055}
\end{equation}

From the observation of the processes (\ref{054}) it follows that
two flavor neutrinos that contribute to $n_{\nu}$
in Eq.~(\ref{055}) are $\nu_{\mu}$
and $\nu_{e}$. The most plausible
candidate for
the third neutrino is $\nu_{\tau}$, discovered in CC reactions.
It is interesting that we still have no direct proof of that (for such a
proof the investigation of  $\nu_{\tau}$-induced NC processes is required).

\section{Lepton number and neutrino mixing}
\label{Lepton number and neutrino mixing}

In the case of neutrino mixing,
the possible existence of a conserved lepton number
can be connected only with neutrinos with definite masses.
The neutrino mass term has the form
\begin{equation}
\mathcal{L}_{\mathrm{mass}}
=
-
\sum_i m_i \, \overline{\nu_{iR}} \, \nu_{iL} + \text{h.c.}
\,.
\label{061}
\end{equation}
If the right-handed components $\nu_{iR}$ and the left-handed components
$\nu_{iL}$ are \emph{independent},
the massive neutrinos $\nu_{i}$ are
Dirac particles.
Indeed, in this case the total Lagrangian is invariant
under the global gauge transformation
\begin{equation}
\nu_{iL} \to e^{i\alpha} \nu_{iL}
\,,
\quad
\nu_{iR} \to e^{i\alpha} \nu_{iR}
\,,
\quad
\ell \to e^{i\alpha} \ell
\,,
\label{062}
\end{equation}
where $\alpha $ is an arbitrary constant.
This invariance implies that the lepton number $L$,
which has the same value
for $e^{-}$,  $\mu^{-}$,  $\tau^{-}$ and all $\nu_{i}$'s,
is conserved.
In this case the
quanta of the fields $\nu_{i}$ are
neutrinos with  $L=1$ and antineutrinos with $L=-1$.

On the other hand,
if the right-handed components $\nu_{iR}$ and
the left-handed components $\nu_{iL}$ are not independent,
but connected by the relation
\begin{equation}
\nu_{iR} = (\nu_{iL})^{c} = \mathcal{C} (\overline{\nu_{iL}})^{T}
\label{063}
\end{equation}
($\mathcal{C}$ is the matrix of charge conjugation),
the massive neutrinos $\nu_{i}$ are Majorana particles.
In this case there is no any gauge invariance of the total
Lagrangian\footnote{
In the Majorana case the transformation 
$\nu_{iL} \to e^{i\alpha} \nu_{iL}$  
requires
$\nu_{iR} \to e^{-i\alpha} \nu_{iR}$.
It is obvious that the
mass term (\ref{061}) is not invariant under these transformations.
}
and the neutrino field $ \nu_{i} = \nu_{iL} + \nu_{iR} $
satisfies the Majorana condition
\begin{equation}
\nu_{i} = (\nu_{i})^{c}
\,.
\label{064}
\end{equation}
This condition implies that the
quanta of the field $\nu_{i}$
are \emph{truly neutral} Majorana neutrinos (identical to
antineutrinos).

The problem of the nature of neutrino with definite masses is one of the 
most fundamental problem of the physics of massive and mixed neutrinos
and is
connected with \emph{the origin of neutrino masses and neutrino mixing}.

Dirac neutrino masses can be generated by the standard Higgs
mechanism.
Majorana neutrino masses require a new
mechanism of neutrino mass generation that is beyond the Standard Model.
One of the most popular mechanisms of neutrino mass generation 
is the see-saw mechanism
\cite{Gell-Mann-Ramond-Slansky-79,Yanagida-79,Mohapatra-Senjanovic-80}.
This mechanism is based on the assumption that
the law of conservation of lepton number
is violated
at a scale that is much larger then the scale of violation of the
electroweak symmetry. The see-saw mechanism allows to connect
the smallness of neutrino masses with a large physical scale that
characterizes the violation of the lepton number conservation law.

In order to reveal the
Dirac or Majorana nature of neutrinos it is necessary to study neutrino
mass effects\footnote{
Dirac neutrinos and Majorana neutrinos are different only if
neutrino masses are different from zero. In the case of standard
electroweak interactions with left-handed neutrino fields there is no
physical difference between massless Dirac and massless Majorana
neutrinos \cite{Marshak-Riazuddin-Ryan-69}.
}.

It is impossible to distinguish massive Dirac and Majorana neutrinos
through the investigation of neutrino oscillations
\cite{Bilenky-Hosek-Petcov-PLB94-80,%
Doi-CP-81,%
Langacker-Petcov-SteigmanToshev-NPB282-88}.
Indeed,
in the case of neutrino mixing the leptonic CC current has the form
\begin{equation}
j_{\alpha}^{\mathrm{CC}}
=
2
\sum_{\ell,i}
\overline{\ell_L} \, \gamma_{\alpha} \, U_{\ell i} \, \nu_{iL}
\,.
\label{065}
\end{equation}
If $\nu_{i}$ are Dirac fields, the
mixing matrix $U$ is determined up to the transformation
\begin{equation}
U_{\ell i} \to e^{-i\alpha_{\ell}} \, U_{\ell i} \, e^{i\beta_{i}}
\,,
\label{066}
\end{equation}
where
$\alpha_{\ell}$ and $\beta_{i}$ are arbitrary parameters.
This is due to the fact that the phases of Dirac fields are arbitrary.

In the
case of Majorana neutrinos,
the Majorana condition (\ref{064}) does not
allow to include arbitrary phases into the fields.
Thus, in the Majorana
case the mixing matrix $U$ is determined only up to the transformation
\begin{equation}
U_{\ell i} \to e^{-i\alpha_{\ell}} \, U_{\ell i}
\,,
\label{067}
\end{equation}
From Eqs.~(\ref{066}) and (\ref{067}) it follows that the number of physical
CP-violating phases in the Dirac and Majorana cases are different\footnote{
For $n$ families the number of physical phases
in the case of Dirac neutrinos is $(n-1)(n-2)/2$
(in this case the number of phases is the same as in the quark case).
In the case of Majorana neutrinos the number of the physical phases is
larger: $n(n-1)/2$,
\textit{i.e.} there are $n-1$ additional phases.
}.

From Eq.~(\ref{021}) one can see that under both transformation
(\ref{066}) and (\ref{067})
the amplitude of $\nu_{\ell} \to \nu_{\ell'}$
transitions is transformed as
\begin{equation}
\mathcal{A}_{\nu_{\ell}\to\nu_{\ell'}}
\to
e^{-i(\alpha_{\ell'}-\alpha_{\ell})}
\,
\mathcal{A}_{\nu_{\ell}\to\nu_{\ell'}}
\,,
\label{068}
\end{equation}
and the probability of
$\nu_{\ell} \to \nu_{\ell'}$
transitions
is invariant under the transformation (\ref{068}).
This means that the transition probability
is independent from the additional phases
in the Majorana case.

The most promising process that allows to investigate the nature
of massive neutrino (Dirac or Majorana?) is 
neutrinoless double-$\beta$ decay of even-even nuclei:
\begin{equation}
\mathcal{N}(A,Z) \to \mathcal{N}(A,Z+2) + e^{-} + e^{-}
\,.
\label{069}
\end{equation}

\begin{figure}[p!]
\begin{center}
\input{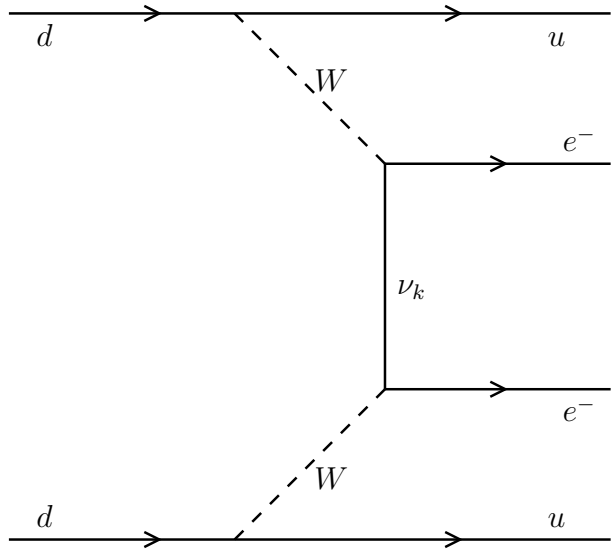}
\end{center}
\caption{ \label{bb}
Diagram of neutrinoless double-$\beta$ decay.
}
\end{figure}

The diagram of this process is depicted
in Fig.~\ref{bb}.
In the case of mixing of Majorana
neutrinos, the neutrino propagator in Fig.~\ref{bb} is given by
\begin{align}
\langle 0 | T \left[ \nu_{eL}(x_1) \nu^T_{eL}(x_2) \right] | 0 \rangle
=
\null & \null
- \sum_i U_{ei}^2
\,
\frac{1-\gamma_5}{2}
\,
\langle 0 | T \left[ \nu_i(x_1) \bar\nu_i(x_2) \right] | 0 \rangle
\,
\frac{1-\gamma_5}{2}
\,
\mathcal{C}
\nonumber
\\
=
\null & \null
- \frac{i}{(2\pi)^4}
\int \mathrm{d}^4p
\,
e^{-i p(x_1-x_2)}
\sum_i U^2_{ei} \, \frac{m_i}{p^2 - m^2_i}
\,
\frac{1-\gamma_5}{2} \, \mathcal{C}
\,.
\label{070}
\end{align}
From Eq.~(\ref{070}) it follows that in the case of small neutrino masses 
the matrix element of neutrinoless double-$\beta$ decay is proportional to
\begin{equation}
\langle{m}\rangle
=
\sum_i U_{ei}^2 \, m_i
\,.
\label{071}
\end{equation}
Thus, the process (\ref{069}) is allowed if
neutrinos are Majorana particles and massive. Notice that neutrino mixing is
not required for that\footnote{
The probability of neutrinoless double-$\beta$ decay is
suppressed because of the
smallness of neutrino masses.
If there is no mixing (\textit{i.e.} $U$ is the unit matrix),
$\langle{m}\rangle = m$, where $m$ is the Majorana mass of $\nu_{e}$.
In neutron decay both
right-handed and left-handed Majorana electron neutrinos are emitted
together with $e^{-}$'s.
The amplitude of the production of left-handed neutrinos is
proportional to $m/E$ ($E$ is the neutrino energy). The left-handed
Majorana neutrino can be absorbed by
another neutron in a nucleus with the production of another $e^{-}$.
The amplitude of absorption of right-handed neutrinos is
proportional to $m/E$.
Hence, the amplitude of neutrinoless double
$\beta$-decay
is proportional to $(m/E) \lesssim 10^{-7}$ for typical nuclear energies
$E \sim 10 \, \mathrm{MeV}$.
}.

Neutrinoless double-$\beta$ 
is a process of second order in the Fermi constant $G_{F}$.
Its matrix element is proportional to small neutrino masses.
The expected lifetime of neutrinoless double-$\beta$ decay  
is much larger than the lifetime of usual $\beta$-decays.  
However, because of the clear signature of the process
(two electrons with definite total energy in the final state),
several experiments have obtained very large lower bounds
for the lifetime of neutrinoless double-$\beta$ decay  
of different nuclei
(see \cite{Tretyak-Zdesenko-95}).
The lower limits for the lifetimes of
$^{76}$Ge and $^{136}$Xe
obtained in the Heidelberg-Moscow experiment
\cite{Heidelberg-Moscow-99}
and in the Gotthard experiment \cite{Gotthard-98}
are
\begin{equation}
T({}^{76}\mathrm{Ge})
\geq
5.7 \times 10^{25} \, \mathrm{yr}
\,,
\quad
T({}^{136}\mathrm{Xe})
\geq
4.4 \times 10^{23} \, \mathrm{yr}
\,,
\quad
\text{at 90\% CL.}
\label{072}
\end{equation}
These limits imply upper bounds for the effective Majorana mass 
$\langle{m}\rangle$,
with values that depend
on the calculation of nuclear matrix elements.
The results of different
calculations lead to the limits
\begin{equation}
\langle{m}\rangle
\leq
0.2-0.6 \, \mathrm{eV}
\,
({}^{76}\mathrm{Ge})
\,,
\quad
\langle{m}\rangle
\leq
2.2-5.2 \, \mathrm{eV}
\,
({}^{136}\mathrm{Xe})
\,,
\quad
\text{at 90\% CL.}
\label{073}
\end{equation}
Several new experiments searching for 
neutrinoless double-$\beta$ decay
are in preparation.
These future experiments are planned to be sensitive to values of
$\langle{m}\rangle$
$\sim 10^{-1} \, \mathrm{eV}$
\cite{CUORE-00,NEMO-3},
or even
$\sim 10^{-2} \, \mathrm{eV}$
\cite{GENIUS-98,BOREXINO-bb-00,Danilov:2000pp}.

The results of neutrino oscillation experiments give
information on neutrino masses
and on the elements of the neutrino mixing matrix.
The possible values of $|\langle{m}\rangle|$,
depend, among others,
on the character of the neutrino mass spectrum,
on the real existence of the oscillations observed in the LSND experiment,
and
on the absolute values of neutrino masses
(see \cite{BGKP-96,BGKM-bb-98,Giunti-Neutrinoless-99,BGGKP-bb-99,%
Klapdor-Pas-Smirnov-00,Bilenky:2001rz}).

If the results of the LSND experiment will not be confirmed by future
experiments,
all the other neutrino data can be explained by the existence
of only three massive and mixed neutrinos.

If there is a hierarchy of neutrino masses,
\begin{equation}
m_{1} \ll m_{2} \ll m_{3}
\,,
\label{074}
\end{equation}
there is a stringent upper bound for the effective Majorana mass
\cite{BGKM-bb-98,BGGKP-bb-99}:
\begin{equation}
|\langle{m}\rangle| \lesssim 10^{-2} \, \mathrm{eV}
\,.
\label{075}
\end{equation}

In the case of an ``inverted hierarchy'',
\begin{equation}
m_{1} \ll m_{2} < m_{3}
\,,
\qquad
m_{1} \ll 1 \, \mathrm{eV}
\,
\label{076}
\end{equation}
the upper bound for the effective Majorana mass is less stringent
\cite{BGGKP-bb-99}:
\begin{equation}
|\langle{m}\rangle| \lesssim 7 \times 10^{-2} \, \mathrm{eV}
\,.
\label{077}
\end{equation}

If the evidence in favor of
short-baseline
$\nu_\mu\to\nu_e$
oscillations
obtained in the LSND experiment will be confirmed by other experiments,
the effective Majorana mass could be as large as
$\sim 1 \, \mathrm{eV}$
or smaller than
$\sim 10^{-2} \, \mathrm{eV}$,
depending on the neutrino mass spectrum
\cite{BGKP-96,BGKM-bb-98,Giunti-Neutrinoless-99,BGGKP-bb-99,%
Klapdor-Pas-Smirnov-00}.

\section{Conclusions}
\label{Conclusions}

We have discussed the notion of lepton numbers 
in the case of neutrino mixing.
We have stressed that the existence of
neutrino oscillations means that there are no conserved family
lepton numbers $L_{e}$, $L_{\mu}$, $L_{\tau}$.
The flavor neutrinos $\nu_{e}$, $\nu_{\mu}$ and $\nu_{\tau}$ 
participate in weak
interactions
($\nu_{\mu}$ is produced together with
a $\mu^{+}$ in $\pi^{+}$ decay, etc.).
In the case of \emph{small} neutrino masses the states 
of flavor neutrinos are superpositions of the states of
neutrinos $\nu_{i}$ with definite masses.
Flavor neutrinos are
not quanta of any field and they have no definite masses.

We have discussed the difference between different flavor neutrino
discovery experiments.
The results of $\nu_\mu\to\nu_\tau$ and $\nu_e\to\nu_\tau$
oscillation experiments clearly demonstrate that $\nu_{\tau}$ is a
new type of flavor neutrino, different from  $\nu_{e}$ and  $\nu_{\mu}$.
The $\nu_{\tau}$ has been detected directly in the recent
DONUT experiment.

We have stressed that a conserved lepton number $L$ can exist only if
massive neutrinos are Dirac particles.
In this case the electron, muon, tau-lepton and
massive neutrinos $\nu_{i}$
have the same values of $L$.
The lepton number
$L$ distinguishes neutrinos from antineutrinos.
Different neutrinos
differ by the value of their masses.

If massive neutrinos are Majorana particles there are no conserved
lepton numbers.
The search for neutrinoless double-$\beta$ decay
is the most promising method to
test the conservation of lepton number in the case of
neutrino mixing.

\begin{flushleft}
\large\textbf{Acknowledgments}
\end{flushleft}
We would like to thank W. Grimus for useful comments
on a preliminary version of this paper.
We would also like to thank
W. Alberico and A. Bottino for useful discussions.

%\bibliography{lep}

\end{document}